\documentclass[prb,aps,amsmath,amssymb,amsfonts,floatfix,superscriptaddress,showpacs,twocolumn]{revtex4-1}
\usepackage{graphicx}
\usepackage{color}

\renewcommand{\i}{\ensuremath{\mathrm{i}}}
\renewcommand{\d}{\ensuremath{\mathrm{d}}}
\newcommand{\bq}{\ensuremath{\mathbf{q}}}
\newcommand{\bk}{\ensuremath{\mathbf{k}}}
\newcommand{\myexp}[1]{\ensuremath{e^{#1}}}

\newcommand{\Ttau}{{\ensuremath{\cal T}}}

\newcommand{\up}{\ensuremath{\uparrow}}
\newcommand{\down}{\ensuremath{\downarrow}}

\newcommand{\moy}[1]{\ensuremath{\left\langle #1 \right\rangle}}

\begin{document}
\title{Orthogonal Polynomial Representation of Imaginary-Time Green's Functions}
\author{Lewin Boehnke}
\affiliation{I. Institut f{\"u}r Theoretische Physik, Universit{\"a}t Hamburg, 
D-20355 Hamburg, Germany}
\author{Hartmut Hafermann}
\affiliation{Centre de Physique Th\'eorique, Ecole Polytechnique, CNRS, 91128 
Palaiseau Cedex, France}
\author{Michel Ferrero}
\affiliation{Centre de Physique Th\'eorique, Ecole Polytechnique, CNRS, 91128 
Palaiseau Cedex, France}
\author{Frank Lechermann}
\affiliation{I. Institut f{\"u}r Theoretische Physik, Universit{\"a}t Hamburg, 
D-20355 Hamburg, Germany}
\author{Olivier Parcollet}
\affiliation{Institut de Physique Th\'eorique (IPhT), CEA, CNRS, URA 2306, 91191
Gif-sur-Yvette, France}

\begin{abstract}
We study the expansion of single-particle and two-particle 
imaginary-time Matsubara Green's functions of quantum impurity models
in the basis of Legendre orthogonal polynomials.
We discuss various applications within the dynamical mean-field theory (DMFT) framework.
The method provides a more compact representation of the Green's functions 
than standard Matsubara frequencies and therefore significantly reduces the memory-storage size of these quantities. 
Moreover, it can be used as an efficient noise filter for various physical quantities 
within the continuous-time quantum Monte Carlo impurity solvers
recently developed for DMFT and its extensions. 
In particular, we show how to use it 
for the computation of energies in the context of realistic DMFT calculations
in combination with the local density approximation to the density functional theory (LDA+DMFT) and for the calculation of lattice 
susceptibilities from the local irreducible vertex function.
\end{abstract}

\maketitle

In recent years, significant progress has been made in the study of strongly-correlated fermionic quantum systems with the development of methods combining systematic analytical approximations and modern numerical algorithms. The Dynamical Mean-Field Theory (DMFT) (for a review see Ref.~\onlinecite{Georges96}) and its various extensions.~\cite{MaierReview,toschi,dualfermion1,clusterdualfermion,ldfa} serve as successful examples for this theoretical advance.
On the technical side, important progress was made in the solution of quantum impurity problems, i.e. local quantum systems coupled to a bath (self-consistently determined in the DMFT formalism). 
In particular, a new generation of continuous-time quantum Monte Carlo (CTQMC) impurity solvers~\cite{Rubtsov05,Werner06,WernerPRBLong,CTAUX}
has emerged that provide unprecedented efficiency and accuracy (for a recent review, see Ref.~\onlinecite{ctqmcrmp}).

In practice, several important technical issues still remain.
Firstly, while the original DMFT formalism is expressed in terms of single-particle quantities (Green's function and self-energy),
two-particle quantities play a central role in the formulation of some DMFT extensions
(e.g. dual-fermions~\cite{dualfermion1,dualfermion2,gangli,brener}, D$\Gamma$A~\cite{toschi}) and in susceptibility and transport computations in DMFT itself.
They typically depend on three independent times or frequencies, and spatial indices. Therefore, they are quite large objects that are hard to store, manipulate and analyze, even with modern computing capabilities.
Developing more compact representations of these objects and using them to solve, e.g., the Bethe-Salpeter equations is therefore an important challenge.

A natural route is to use an orthogonal polynomial representation of the imaginary-time dependence of these objects. While the application of orthogonal polynomials has had productive use in other approaches to correlated electrons,~\cite{KPM,CheMPS} in this paper we show how to use Legendre polynomials to represent various imaginary-time Green's functions in a more compact way and show their usefulness in some concrete calculations. 

A second aspect is that modern CTQMC impurity solvers still have limitations. One well-known problem is the high-frequency noise observed in the Green's function and the self-energy (see e.g. Fig.~6 of Ref.~\onlinecite{Gull07}).
Even though this is in general of little concern for the DMFT self-consistency itself, 
it can become problematic when computing the energy, since the precision depends crucially on the high-frequency expansion coefficients of the Green's function and self-energy.
An important field of application involves realistic models of strongly correlated materials through the combination with the local density approximation (LDA+DMFT).~\cite{KotliarRMP06}
In this paper, we show that physical quantities such as the Green's function, kinetic energy, and even the coefficients of the high-frequency expansion of the Green's function can be measured directly in the Legendre representation within CTQMC and that the basis truncation acts as a very efficient noise filter: the statistical noise is mostly carried by high-order Legendre coefficients, while the physical properties are determined by the low-order coefficients.

This paper is organized as follows: Section I is devoted to single-particle
Green's functions. More precisely, in Sec. I.A, we introduce the Legendre
representation of the single-particle Green's function and how it appears in
the CTQMC context; we then illustrate the method on the imaginary-time (I.B)
and imaginary-frequency (I.C) Green's function of a standard DMFT computation;
in I.D, we discuss the use of the Legendre representation to compute the
energy in a realistic computation for SrVO$_{3}$.  Section II is
devoted to two-particle Green's functions: We first present the expansion in
Sec. II.A and illustrate it on an explicit DMFT computation of the antiferromagnetic
susceptibility in Sec. II.B, followed by the example of a calculation of the dynamical
wave-vector resolved magnetic susceptibility.  Additional information can be found
in the appendixes. Appendix~\ref{sec:propertieslegendre} gives some properties of
the Legendre polynomials relevant for this work. 
Appendix~\ref{sec:Rapiddecayproof} discusses the rapid decay of the
Legendre coefficients of the single-particle Green's function.
Appendix~\ref{App.GAccu} first
derives the accumulation formulas for the single-particle and two-particle
Green's functions in the hybridization expansion CTQMC (CT-HYB)
algorithm~\cite{Werner06} (while these formulas have been given before,~\cite{Werner06,Gull07}
the proof presented here aims to explain their resemblance to a Wick's theorem).
We then give the explicit formulas in the Legendre basis. For completeness, we
provide an accumulation formula for the continuous- time interaction expansion
(CT-INT)\cite{Rubtsov05} and auxiliary field (CT-AUX)\cite{CTAUX} algorithms in
Appendix \ref{app:CTINTacc}. Finally, 
in Appendix~\ref{sec:Tnulformula}, we derive the expression for the matrix that
relates the coefficients of the Green's function in the Legendre representation
to its Matsubara frequency representation.

\section{Single-particle Green's function}

\subsection{Legendre representation}

We consider the single-particle imaginary-time Green's function $G(\tau)$ defined 
on the interval $[0,\beta]$, where
$\beta$ is the inverse temperature. 
Expanding $G(\tau)$ in terms of Legendre polynomials $P_l(x)$ defined on the interval $[-1,1]$, we have
\begin{align}
  G(\tau) =& \sum_{l\geq0}\frac{\sqrt{2l+1}}{\beta} P_l(x(\tau)) \, G_l,
    \label{eqn:gtauleg} \\
  G_l     =& \sqrt{2l+1} \int_0^\beta d\tau \, P_l(x(\tau)) \, G(\tau).
    \label{eqn:G2leg}
\end{align}
where $x(\tau) = 2\tau/\beta -1$ and $G_{l}$ denote the coefficients of $G(\tau)$ in the Legendre basis. The most important properties of the Legendre polynomials are summarized in Appendix~\ref{sec:propertieslegendre}. 

We note that a priori different orthogonal polynomial bases (e.g., Chebyshev
instead of Legendre polynomials) may be used, and many of the conclusions in
this paper would remain valid. The advantage of the Legendre polynomials is
that the transformation between the Legendre representation and the Matsubara
representation can be written in terms of a unitary matrix, since Legendre
polynomials are orthogonal with respect to a scalar product that does not
involve a weight function (see below and Appendix \ref{sec:Tnulformula}).
In this paper, therefore we restrict our discussion to the Legendre polynomials.

On general grounds, one can expect the Legendre representation of $G(\tau)$ to be {\sl much more compact than the standard Matsubara representation}: in order to perform a Fourier series expansion in terms of Matsubara frequencies, $G(\tau)$ has to be anti-periodized for all $\tau\in \mathbb{R}$, while the full information is already contained in the interval $[0,\beta]$. As a result, the Green's function contains discontinuities in $\tau$ that result in a slow decay at large frequencies (typically $\sim 1/\nu_n$).
On the other hand, expanding $G(\tau)$, which is a smooth function of $\tau$
on the interval $[0,\beta]$, in terms of Legendre polynomials yields coefficients $G_{l}$ that decay faster than the inverse of any power of $l$ (as shown in Appendix \ref{sec:Rapiddecayproof}). As a result, the information about a Green's function can be saved in a very small storage volume. As we will show in Sec.~\ref{sec:2part}, this is particularly relevant when dealing with more complex objects such as the two-particle Green's function, which depends on three frequencies.

CTQMC algorithms usually measure the Green's function $G(\tau)$ in one of the
two following ways: (i) using a very fine grid for the interval
$[0,\beta]$ or (ii) measuring the Fourier transform of the Green's
function on a finite set of Matsubara frequencies.~\cite{Rubtsov05,Haule07}
We show in Appendix~\ref{sec:accumulation} explicitely for the
CT-HYB~\cite{Werner06,ctqmcrmp} algorithm, that one can also directly measure the
coefficients $G_l$ during the Monte Carlo process
(we expect our conclusions to hold for any continuous-time Monte Carlo algorithm).

%~~~~~~~~~~~~~~~~~~~~~~~~~~~~~~~~~~~~~~~~~~~~~~~~~~~~~~~~~~~~~~~~~~~~~~~~~~~~~~~
\begin{figure}[!Htb]
\begin{center}
  \includegraphics[scale=.65,angle=0]{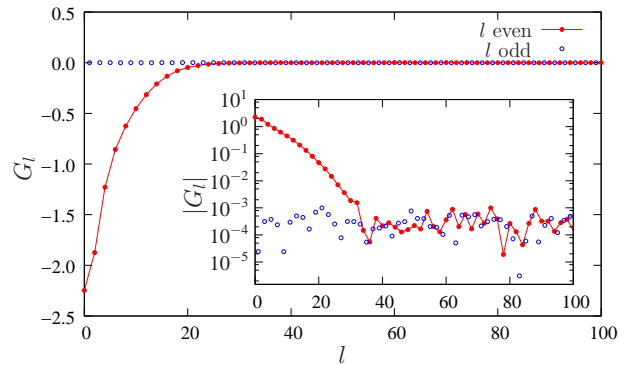}
\end{center}
\caption{(Color online) Legendre coefficients $G_l$ of the Green's function of the half-filled 
         Hubbard model on the Bethe lattice within DMFT. Error bars are not shown on the logarithmic plot. They are of the order of $10^{-4}$.}
\label{fig:G2lt}
\end{figure}
%~~~~~~~~~~~~~~~~~~~~~~~~~~~~~~~~~~~~~~~~~~~~~~~~~~~~~~~~~~~~~~~~~~~~~~~~~~~~~~~

As an illustration, we will focus on the
Green's function obtained by DMFT for the Hubbard
model at half-filling described by the Hamiltonian
\begin{equation}\label{eqn:Hubbard}
  H = -t \sum_{\langle ij\rangle\sigma} c_{i\sigma}^{\dagger}c_{j\sigma}
              + U \sum_{i}n_{i\uparrow}\,n_{i\downarrow},
\end{equation}
where $c_{i\sigma}^{(\dagger)}$ creates (annihilates) an electron with spin $\sigma$ on the site $i$ of a Bethe lattice~\cite{bethe,Georges96} and $\langle ij\rangle$ on the sum denotes nearest neighbors. In the following, quantities will be
expressed in units of the hopping $t$ and we set the on-site Coulomb repulsion to $U/t=4$ and use the temperature $T/t=1/45$.
We solve the DMFT equations using the TRIQS\cite{TRIQS} toolkit and its 
implementation of the CT-HYB~\cite{Werner06,ctqmcrmp} algorithm.
 In Fig.~\ref{fig:G2lt}, we show the coefficients $G_l$ that we obtain. 
Note that coefficients for $l$ odd must be zero due to particle-hole symmetry.
Indeed, the coefficients in our data for odd $l$'s all take on very small value, compatible with a vanishing value within their error bars. 
The even $l$ coefficients instead show a very fast decay, as discussed above.
For $l > 30$, all coefficients eventually take values of the order of the
statistical error bar.

Let us now discuss the specific issue of the statistical Monte Carlo noise.
We observe that the high-order Legendre coefficients 
have a larger relative noise than small $l$ coefficients.
On general grounds, we expect the coefficients of the {\it exact} Green's function to continue to decrease faster than any power of $1/l$ to zero (cf. Appendix~\ref{sec:Rapiddecayproof}). 
Hence, physical quantities computed from $G(\tau)$ are likely to have a
very weak dependence on the $G_l$ for large $l$. A good approximation then is to truncate the expansion in Legendre polynomials at an order $l_\mathrm{max}$ and set $G_l = 0$ for $l > l_\mathrm{max}$. The choice for $l_\mathrm{max}$ has to be such that the quantity of interest is accurately represented. On the other hand, if $l_\mathrm{max}$ is too large, we would start to include coefficients that have increasingly large error bars compared to their value and this would eventually pollute the calculation. 
A systematic method is therefore to examine the physical quantity as a function of the cutoff $l_\mathrm{max}$.
We expect that it will first reach a plateau where it is well converged. 
The existence of a plateau means that the contribution of higher-order coefficients is indeed negligible. For larger $l_\mathrm{max}$, the statistical noise in the $G_l$ will destabilize this plateau whose size will increase with the precision of the CTQMC computation. The existence of such a plateau provides a controlled way to determine the adequate value of $l_\mathrm{max}$.
In the remaining paragraphs of this section, we will illustrate this phenomenon 
on different physical quantities by studying
their dependence on $l_\mathrm{max}$.

\subsection{Imaginary-time Green's function}

%~~~~~~~~~~~~~~~~~~~~~~~~~~~~~~~~~~~~~~~~~~~~~~~~~~~~~~~~~~~~~~~~~~~~~~~~~~~~~~~
\begin{figure}[!Htb]
\begin{center}
  \includegraphics[scale=.65,angle=0]{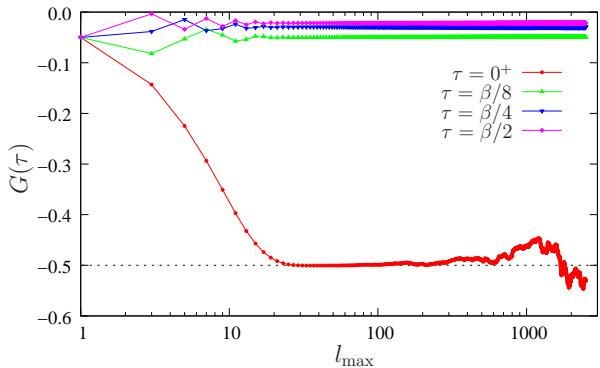}
\end{center}
\caption{(Color online) Imaginary-time Green's function $G(\tau)$ at four different values of
         $\tau$ as a function of $l_\mathrm{max}$.}
\label{fig:Gtvslmax}
\end{figure}
%~~~~~~~~~~~~~~~~~~~~~~~~~~~~~~~~~~~~~~~~~~~~~~~~~~~~~~~~~~~~~~~~~~~~~~~~~~~~~~~

It is instructive to analyze the effect of $l_\mathrm{max}$ on the
reconstructed imaginary-time Green's function $G(\tau)$ (using
Eq.~\eqref{eqn:gtauleg}). In Fig.~\ref{fig:Gtvslmax}, we show the evolution of
$G(\tau)$ at $\tau = 0^+$, $\tau = \beta/8$, $\tau = \beta/4$, and $\tau = \beta/2$ with the cutoff. It is apparent that these values very rapidly converge as a function of $l_\mathrm{max}$. We observe a well-defined and extended plateau.
As the cutoff grows bigger, noise reappears in $G(\tau)$ because of the comparatively large error bars in higher-order $G_l$'s.

%~~~~~~~~~~~~~~~~~~~~~~~~~~~~~~~~~~~~~~~~~~~~~~~~~~~~~~~~~~~~~~~~~~~~~~~~~~~~~~~
\begin{figure}[!Htb]
\begin{center}
  \includegraphics[scale=.65,angle=0]{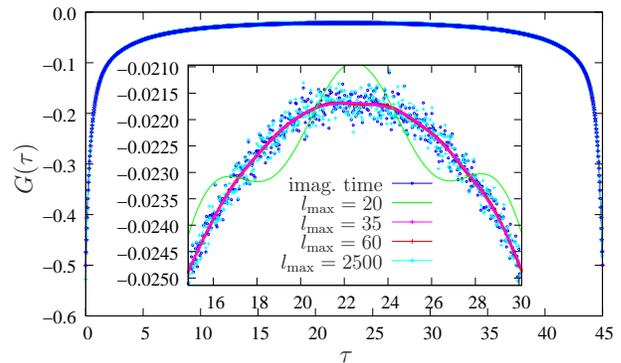}
\end{center}
\caption{(Color online) Imaginary-time Green's function $G(\tau)$ on the interval $[0,\beta]$
         measured on a finite 1500-bin mesh (blue scattered points) and computed from
         $l_\mathrm{max}$ Legendre coefficients (solid lines). Four different choices for
         $l_\mathrm{max}$ are shown.  Inset: zoom on the area around $\beta/2$.}
\label{fig:Gt}
\end{figure}
%~~~~~~~~~~~~~~~~~~~~~~~~~~~~~~~~~~~~~~~~~~~~~~~~~~~~~~~~~~~~~~~~~~~~~~~~~~~~~~~

In Fig.~\ref{fig:Gt}, the Green's function is reconstructed on the full interval $[0,\beta]$ and compared to a direct measurement on a 1500-bin mesh. For $l_\mathrm{max} = 20$, where the individual values of $G(\tau)$ have not yet converged to their plateau (see Fig.~\ref{fig:Gtvslmax}), the resulting Green's function is smooth but not compatible with the scattered direct measurements. For $l_\mathrm{max} = 35$ and $60$, $G(\tau)$ is smooth and nicely interpolates the scattered data. Moreover $G(\tau)$ is virtually identical for both values of $l_\mathrm{max}$. This is expected because both of these values lie on the plateau. When $l_\mathrm{max}$ is very large, i.e., of the order of the number of imaginary-time bins, the noise in $G(\tau)$ eventually reappears and begins to resemble that of the direct measurement. 
We emphasize that all measurements have been performed within the same calculation and hence contain identical statistics. Hence the information in both measurements is identical up to the error committed by truncating the basis.

It is clear from this analysis, that the truncation of the Legendre basis acts as a \emph{noise filter}. We note that no information is lost by the truncation:
the high-order coefficients correspond to information on very fine details of the Green's function, which cannot be resolved within a Monte Carlo calculation, as is obvious from the noisy $G(\tau)$. 

\subsection{Matsubara Green's function and high-frequency expansion}

It is common to use the Fourier transform $G(\i\nu_n)$ of $G(\tau)$ to
manipulate Green's functions. This representation is, for example, convenient to
compute the self-energy from Dyson's equation or to compute correlation
energies.
In terms of $G_l$, we can obtain the Matsubara Green's function with
\begin{align}
  G(\i\nu_n) &= \sum_{l\geq0}G_l\frac{\sqrt{2l+1}}{\beta}\!\int_0^\beta\!\!\!d\tau\,
  \operatorname{e}^{\i\nu_n\tau} P_l(x(\tau))\nonumber\\
  &= \sum_{l\geq0}T_{n l}G_l.
  \label{eqn:Tnul}
\end{align}
where the unitary transformation $T_{nl}$ is shown in 
Appendix~\ref{sec:Tnulformula} to be
\begin{equation}
  T_{nl}=  (-1)^n\,\i^{l+1} \sqrt{2l+1}\,\, j_l\left(\frac{(2n+1)\pi}{2}\right)
\end{equation}
with $j_l(z)$ denoting the spherical Bessel functions. Note that $T_{nl}$ is independent of $\beta$.

%~~~~~~~~~~~~~~~~~~~~~~~~~~~~~~~~~~~~~~~~~~~~~~~~~~~~~~~~~~~~~~~~~~~~~~~~~~~~~~~
\begin{figure}[!Htb]
\begin{center}
  \includegraphics[scale=.65,angle=0]{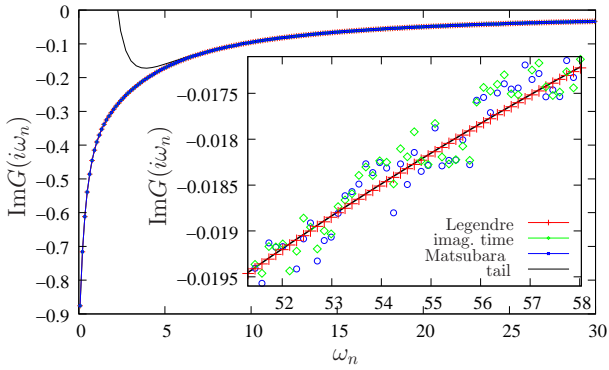} 
\end{center}
\caption{(Color online) Matsubara Green's function obtained from measurements
         made directly on the Matsubara frequencies (blue scattered
         points), calculated from an imaginary-time measurement (green scattered points) and computed from
         Eq.~\eqref{eqn:Tnul} with $l_\mathrm{max} = 35$ (red solid line).
         The analytically known high-frequency tail is shown for comparison
         (black solid line).  Inset: Blowup of the high-frequency region.}
\label{fig:gw}
\end{figure}
%~~~~~~~~~~~~~~~~~~~~~~~~~~~~~~~~~~~~~~~~~~~~~~~~~~~~~~~~~~~~~~~~~~~~~~~~~~~~~~~

In Fig.~\ref{fig:gw}, we display the Matsubara Green's function as measured
directly on the Matsubara axis and as computed from Eq.~\eqref{eqn:Tnul} with a fixed cutoff $l_\mathrm{max}$. The direct measurement of $G(\i\nu_n)$ has been done
within the same Monte Carlo simulation as the one used to compute the $G_l$
discussed above. It is clear from the plot that the truncation to $l_\mathrm{max}$
has filtered the high-frequency noise, and that for large $\i\nu_n$ the
Matsubara Green's function has a smooth power-law decay. Let us emphasize here that the
Matsubara Green's function is obtained in an unbiased manner that does not involve
any model-guided Fourier transform (see also Ref.~\onlinecite{gunnarsson}).

We will now show that the coefficients that control this power-law decay can also be
accurately computed. Let us consider the high-frequency
expansion of $G(\i\nu_n)$
\begin{equation}
  G(\i\nu_n)=\frac{c_1}{\i\nu_n}+\frac{c_2}{(\i\nu_n)^2}+\frac{c_3}{(\i\nu_n)^3}+\dots
\label{eqn:moments}
\end{equation}
Using the known high-frequency expansion of $T_{nl}$ (cf. Appendix~\ref{sec:Tnulformula}),
\begin{equation}
T_{nl}=\frac{t^{(1)}_{l}}{\i\nu_n\beta }+\frac{t^{(2)}_{l}}{ (\i\nu_n\beta)^2}+\frac{t^{(3)}_{l}}{ (\i\nu_n\beta)^3}+\dots\label{eqn:tmoments},
\end{equation}
one can directly relate the $c_p$ and the $G_l$. Indeed,
from \eqref{eqn:Tnul}, \eqref{eqn:moments}, and \eqref{eqn:tmoments}, it follows that
\begin{equation}
c_p=\frac{1}{\beta^p}\sum_{l\geq0} t^{(p)}_lG_l.
\end{equation}
The general expression of the coefficients $t^{(p)}_l$ is shown in~\eqref{eq:smallt}.
For the first three moments, we have the following expressions
\begin{subequations}
\begin{align}
  c_{1} &= -\sum_{l\geq0,\, \text{even}} \frac{2\sqrt{2l+1}}{\beta}\, G_{l}\label{eq:c1}\\
  c_{2} &= +\sum_{l>0,\, \text{odd}} \frac{2\sqrt{2l+1}}{\beta^{2}}\, G_{l}\, l(l+1)\label{eq:c2}\\
  c_{3} &= -\sum_{l\geq0,\,\text{even}} \frac{\sqrt{2l+1}}{\beta^{3}}\, G_{l}\, (l+2)(l+1)l(l-1)\label{eq:c3}.
\end{align}
\end{subequations}
Since $t^{(p)}_l \sim l^{2p-3/2}$, with the fast decay of the $G_l$ discussed above, 
we can expect a stable convergence of the $c_p$ as a function
of $l_\mathrm{max}$.
Note, however, that  when $p$ increases, 
the coefficients grow, so we expect to need more and more Legendre coefficients
to compute the series in practice.

%~~~~~~~~~~~~~~~~~~~~~~~~~~~~~~~~~~~~~~~~~~~~~~~~~~~~~~~~~~~~~~~~~~~~~~~~~~~~~~~
\begin{figure}[!Htb]
\begin{center}
  \includegraphics[scale=.65,angle=0]{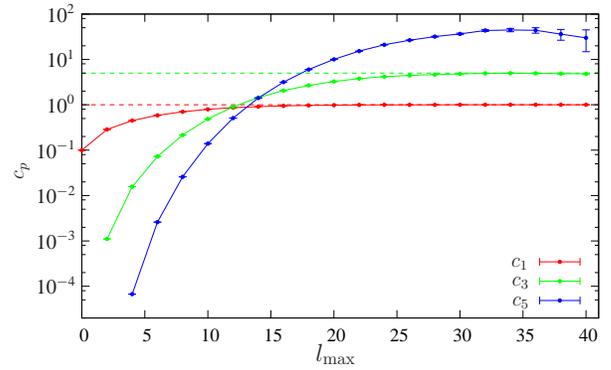} 
\end{center}
\caption{(Color online) Convergence of the moments $c_{1}$, $c_{3}$ and $c_{5}$ as a function of $l_\mathrm{max}$.
         Only points corresponding to an even cutoff are shown because odd terms in the sum vanish.
         The analytically known results for $c_{1}$ and $c_{3}$ are indicated by dashed lines.
         Even moments are zero due to particle-hole symmetry.}
\label{fig:momU4}
\end{figure}
%~~~~~~~~~~~~~~~~~~~~~~~~~~~~~~~~~~~~~~~~~~~~~~~~~~~~~~~~~~~~~~~~~~~~~~~~~~~~~~~

The convergence of the moments is illustrated in Fig.~\ref{fig:momU4}. For the
model we consider, the first moments are explicitly given by
\begin{align}
  c_1 &= 1 \quad c_2 = 0 \nonumber\\
  c_3 &= 5 \quad c_4 = 0. \nonumber
\end{align}
We see that $c_1$ and $c_3$ smoothly converge to a plateau. For the higher
moment $c_5$, a larger number of Legendre coefficients is required. A plateau
is reached but is (depending on the accuracy of the data accumulated in the QMC simulation) quickly destabilized when $l_\mathrm{max}$ gets bigger
and noisy $G_l$ are included in the calculation. This clearly shows that
$l_\mathrm{max}$ has to be chosen carefully to get sensitive $c_p$.
For larger cutoff the error in the moments grows rapidly. 
This shows that a large error on the high-frequency moments is committed when measuring in a basis in which it is not possible to filter the noise, i.e., the conventional imaginary-time or Matsubara representation.

Note that it is easy to incorporate a priori information on the moments $c_p$. For
example, in the model we consider above, we have $c_1 = \langle \{ c, c^\dagger \} \rangle = 1$. From (\ref{eq:c1}), we see that this is a {\it linear}
constraint on the $G_l$ coefficients, which we can therefore enforce by projecting the Legendre coefficients onto the ($l_\mathrm{max}$)-dimensional hyperplane defined by the constraint (\ref{eq:c1}).
A correction to impose, e.g.,  a particular $c_1$ is straightforwardly found to be
\begin{equation}
  G_l \rightarrow G_l + \left( \beta c_1 - \sum_{l^\prime=0}^{l_\mathrm{max}}
  t^{(1)}_{l^\prime} G_{l^\prime} \right) \frac{t^{(1)}_l}{\sum_l|t^{(1)}_l|^2}.
\end{equation}
This is easily generalized to other constraints.

\subsection{Energy}

The accurate determination of the high-frequency coefficients is of central
importance, since many quantities are computed from sums
over all Matsubara frequencies involving $G(\i\nu_n)$. Because $G(\i\nu_n)$ slowly decreases as $\sim 1/(\i\nu_n)$ to leading order, these sums are usually computed from the
actual data up to a given Matsubara frequency and the remaining frequencies are summed
up analytically from the knowledge of the $c_p$.  Thus, an incorrect determination
of the $c_p$ leads to significant numerical errors. This is a particularly delicate
issue when $G(\i\nu_n)$ is measured directly on the Matsubara axis. In this case
one usually needs to fit the noisy high-frequency data to
infer the high-frequency moments. As discussed above, such a procedure is not
required when using Legendre coefficients and the $c_p$ can be computed in a
controlled manner. In the following, we illustrate this point in an actual
energy calculation.

Based on an LDA+DMFT calculation for the compound SrVO$_{3}$,~\cite{amadon2008,aichhorn2009} we compute the kinetic energy $E_\mathrm{kin} = (1/N) \sum_{k,\alpha} \langle n_{k \alpha}\rangle \epsilon_{k \alpha}$ and the
correlation energy $E_\mathrm{corr} = (1/N) \sum_i U \langle n_{i \uparrow}
n_{i \downarrow} \rangle$ ($N$ denotes the number of lattice sites) resulting from the implementation and parameters of
Ref.~\onlinecite{aichhorn2009}. These terms are contributions to the LDA+DMFT total energy~\cite{Amadon06} which depend explicitly on the results of the DMFT impurity solver.

\begin{figure}[!Htb]
\begin{center}
  \includegraphics[scale=.65,angle=0]{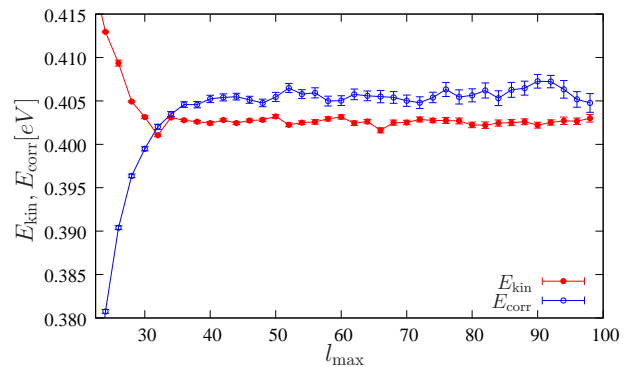} 
\end{center}
\caption{(Color online) Kinetic energy $E_\mathrm{kin}$ (full symbols) and correlation energy
         $E_\mathrm{corr}$ (open symbols) for SrVO$_{3}$ as a function of $l_\mathrm{max}$,
         computed with the implementation and parameters of
         Ref.~\onlinecite{aichhorn2009}. For clarity the kinetic energy has been
         shifted by $384.86 \mathrm{eV}$. Error bars are computed from 80 converged LDA+DMFT iterations.}
\label{fig:energies}
\end{figure}

The results are shown in Fig.~\ref{fig:energies}. Here the parameter $l_\mathrm{max}$, against which these quantities are plotted, represents the number of Legendre coefficients used throughout the LDA+DMFT self-consistency. It is also the number of coefficients used to evaluate $\langle n_{k\alpha}\rangle$ from the lattice Green's function $G_{k}(\i\nu_n)$. Note that $E_\mathrm{corr}$ has been accumulated directly within the CTQMC simulation.
%
%~~~~~~~~~~~~~~~~~~~~~~~~~~~~~~~~~~~~~~~~~~~~~~~~~~~~~~~~~~~~~~~~~~~~~~~~~~~~~~~
%~~~~~~~~~~~~~~~~~~~~~~~~~~~~~~~~~~~~~~~~~~~~~~~~~~~~~~~~~~~~~~~~~~~~~~~~~~~~~~~

In agreement with an analysis of the convergence with respect to the number of Legendre coefficients $l_\mathrm{max}$ similar to the ones shown in Figs. \ref{fig:Gtvslmax}, \ref{fig:momU4} for an \emph{individual} DMFT iteration, we find a plateau for both energies at $l_\mathrm{max} \sim 40$.
While the energy can be accurately computed within a single DMFT iteration, the error here mainly stems from the fluctuations between successive DMFT iterations. The plateau remains up to the largest values of $l_\mathrm{max}$.
However, as $l_\mathrm{max}$ gets larger, so do the error bars, due to the feedback of noise from the largest Legendre coefficients.
Note that the error bars on the correlation energy, computed directly within the CTQMC algorithm, are of the same order of magnitude as those on the kinetic energy.
The existence of a plateau implies that for a well-chosen cutoff $l_\mathrm{max}$, the energy can be computed in a controlled manner.
We want to emphasize that such an approach is simpler and better controlled than delicate fitting procedures of high-frequency tails of the Green's function on the Matsubara axis.

\section{Two-particle Green's function}
\label{sec:2part}

\subsection{Legendre representation for two-particle Green's functions}

The use of Legendre polynomials proves very useful when dealing with
two-particle Green's functions. We will show that it brings about improvements both
from the perspective of storage size and convergence as a function of the
truncation. The object one mainly deals with is the generalized susceptibility,
\begin{align}
  \widetilde{\chi}^{\sigma\sigma^\prime} & (\tau_{12}, \tau_{34}, \tau_{14}) =
  \widetilde{\chi}^{\sigma\sigma^\prime}   (\tau_1 - \tau_2, \tau_3 - \tau_4, \tau_1 - \tau_4) = \nonumber \\
&  \big< \Ttau c^\dagger_\sigma (\tau_1) c_\sigma (\tau_2)
                               c^\dagger_{\sigma^\prime} (\tau_3) c_{\sigma^\prime} (\tau_4) \big> \nonumber \\
&- \big< \Ttau c^\dagger_\sigma (\tau_1) c_\sigma (\tau_2) \big>
   \big< \Ttau c^\dagger_{\sigma^\prime} (\tau_3) c_{\sigma^\prime} (\tau_4) \big>.
\label{eqn:chitau}
\end{align}
Let us emphasize that $\widetilde{\chi}$ is a function of three independent time-differences only.
With the particular choice made above, $\widetilde{\chi}$ is $\beta$-antiperiodic in $\tau_{12}$ and
$\tau_{34}$, while it is $\beta$-periodic in $\tau_{14}$. Consequently, its Fourier
transform $\widetilde{\chi}(\i\nu_n,\i\nu_{n^\prime},\i\omega_m)$ is a function of two
fermionic frequencies $\nu_n = 2(n+1)\pi/\beta$, $\nu_{n^\prime}=2(n^\prime+1)\pi/\beta$, and one
bosonic frequency $\omega_m = 2m\pi/\beta$.

We introduce a representation of $\widetilde{\chi}(\tau_{12}, \tau_{34}, \tau_{14})$ in terms of the coefficients $\widetilde{\chi}_{ll^\prime} (\i\omega_m)$ such that
\begin{align}
  &\widetilde{\chi} (\tau_{12}, \tau_{34}, \tau_{14}) =
    \sum_{l,l^\prime\geq0}\sum_{m\in\mathbb{Z}} \frac{\sqrt{2l+1}\sqrt{2l^\prime+1}}{\beta^3} (-1)^{l^\prime+1} \nonumber \\
  & \quad P_l(x(\tau_{12})) P_{l^\prime}(x(\tau_{34})) \myexp{\i\omega_m \tau_{14}} \widetilde{\chi}_{ll^\prime} (\i\omega_m).
\label{eqn:chitaullw1}
\end{align}
In this mixed basis representation, the $\tau_{12}$  and $\tau_{34}$ dependence of $\widetilde{\chi} (\tau_{12}, \tau_{34}, \tau_{14})$ is expanded in terms of Legendre polynomials, while the $\tau_{14}$ dependence is described through Fourier modes $\myexp{\i\omega_m\tau_{14}}$.  The motivation behind this choice is that many equations involving generalized susceptibilities (like the Bethe-Salpeter equation) are diagonal in $\i\omega_m$.
The inverse of~\eqref{eqn:chitaullw1} reads
\begin{align}
\widetilde{\chi}_{ll^\prime} (\i\omega_m) = &
    \iiint d\tau_{12} d\tau_{34} d\tau_{14} \sqrt{2l+1}\sqrt{2l^\prime+1} (-1)^{l^\prime+1} \nonumber \\
  & \quad P_l(x(\tau_{12})) P_{l^\prime}(x(\tau_{34})) \myexp{-\i\omega_m \tau_{14}} \widetilde{\chi} (\tau_{12}, \tau_{34}, \tau_{14}).
\label{eqn:chitaullw2}
\end{align}
We show in Appendix~\ref{sec:accumulation} how the Legendre expansion coefficients of the one- and two-particle 
Green's function (hence of 
$\widetilde{\chi}_{ll^\prime} (\i\omega_m)$ )
can be measured directly within CT-HYB. With the above definition, the Fourier transform $\widetilde{\chi}(\i\nu_n,\i\nu_{n^\prime},\i\omega_m)$
is easily found with
\begin{equation}
  \widetilde{\chi}(\i\nu_n,\i\nu_{n^\prime},\i\omega_m) = \sum_{l,l\geq0}
    T_{n l} \widetilde{\chi}_{ll^\prime}(\i\omega_m) T^\ast_{{n^\prime} l^\prime}.
\label{eqn:chifouriervschilegendre}
\end{equation}
$T_{n l}$ was already defined in Eq.~\eqref{eqn:Tnul}.  Using the additional unitarity property of
$T$ in Eq.~\eqref{eqn:chifouriervschilegendre} one can in general easily rewrite
equations involving the Fourier coefficients $\widetilde{\chi} (\i\nu_n,
\i\nu_{n^\prime},\i\omega_m)$ in sole terms of the $\widetilde{\chi}_{ll^\prime} (\i\omega_m)$.

In the DMFT framework, the lattice susceptibility
$\widetilde{\chi}_\mathrm{latt}$ is obtained from~\cite{Georges96,MaierReview}
\begin{align}
  \Big[\underline{\underline{\widetilde{\chi}_\mathrm{latt}}}\Big]^{-1}   \!\!\!\!\!(\i\omega_m,\bq)
  = \phantom{-}&\Big[\underline{\underline{\widetilde{\chi}_\mathrm{loc}}}\Big]^{-1}    \!\!\!\!\!(\i\omega_m)\nonumber\\
  - &\Big[\underline{\underline{\widetilde{\chi}^0_\mathrm{loc}}}\Big]^{-1}  \!\!\!\!\!(\i\omega_m)
  + \Big[\underline{\underline{\widetilde{\chi}^0_\mathrm{latt}}}\Big]^{-1} \!\!\!\!\!(\i\omega_m,\bq),
\label{eqn:chiq}
\end{align}
where the double underline emphasizes that this is to be thought of as a matrix equation
for the coefficients $\widetilde{\chi}$ expressed either in $(\i\nu_n,\i\nu_{n^\prime})$ in the Fourier
representation or in $(l,l^\prime)$ in the mixed Legendre-Fourier representation.
The bare susceptibilities are given by
\begin{align}
  &\widetilde{\chi}_\mathrm{loc}^0 (\i\nu_n,\i\nu_{n^\prime},\i\omega_m)
    = - G_\mathrm{loc}(\i\nu_n+\i\omega_m) G_\mathrm{loc}(\i\nu_n) \delta_{n,n^\prime}, \nonumber \\
  &\widetilde{\chi}_\mathrm{latt}^0 (\i\nu_n,\i\nu_{n^\prime},\i\omega_m,\bq)\nonumber\\
    &\qquad\qquad= -\sum_{\bk} G^\mathrm{latt}_{\bk+\bq}(\i\nu_n+\i\omega_m) G^\mathrm{latt}_\bk(\i\nu_n) \delta_{n,n^\prime},
\label{eqn:chi0}
\end{align}
where
\begin{equation}
  G^\mathrm{latt}_\bk(\i\nu_n) = \left[\i\nu_n + \mu -
    \epsilon_\bk - \Sigma_\mathrm{loc}(\i\nu_n) \right]^{-1},
\label{eqn:g}
\end{equation}
and $G_\mathrm{loc}$, $\Sigma_\mathrm{loc}$ are the Green's function and self-energy
of the local DMFT impurity problem, respectively. The equivalent susceptibilities in the mixed
Legendre-Fourier representation are simply obtained as the inverse of
Eq.~\eqref{eqn:chifouriervschilegendre}
\begin{equation}
  \widetilde{\chi}^0_{ll^\prime}(\i\omega_m,\bq) = \sum_{n,n^\prime\in\mathbb{Z}}
    T^\ast_{n l} \widetilde{\chi}^0(\i\nu_n,\i\nu_{n^\prime},\i\omega_m,\bq) T_{{n^\prime} l^\prime},
\end{equation}
where the high-frequency behavior of $G_\mathrm{loc}(\i\nu_n)$ and $G^\mathrm{latt}(\i\nu_n)$ can easily be considered in the frequency sums.
Evaluation of lattice susceptibilities from $\widetilde{\chi}_\mathrm{latt}(\i\nu_n,\i\nu_{n^\prime},\i\omega_m)$ can also directly be propagated to the mixed Legendre-Fourier representation, abolishing altogether the need to transform back to Fourier representation
\begin{align}
  \chi(\i&\omega_m,\bq) = \frac{1}{\beta^2}
    \sum_{nn^\prime\in\mathbb{Z}}\widetilde{\chi}_\mathrm{latt} (\i\nu_n,\i\nu_{n^\prime},\i\omega_m,\bq)\nonumber\\
&= \frac{1}{\beta^2}
    \sum_{ll^\prime\geq0} (-1)^{l+l^\prime}\sqrt{2l+1}\sqrt{2l^\prime+1}
    \widetilde{\chi}_{\mathrm{latt},ll^\prime} (\i\omega_m,\bq).
\label{eqn:chisuml}
\end{align}
Note that $\widetilde\chi_\mathrm{latt}$ can be written as the sum of a free two-particle
propagation $\widetilde\chi^0_\mathrm{latt}$ Eq.~\eqref{eqn:chi0} (bubble part) and a
connected rest $\widetilde\chi^\mathrm{conn}_\mathrm{latt}$ (vertex part).
These two terms can be separately summed in Eq.~\eqref{eqn:chisuml}.

The present mixed basis representation has been successfully used in a recent
investigation of static finite-temperature lattice charge and magnetic
susceptibilities for the Na$_x$CoO$_2$ system at intermediate-to-larger doping
$x$.~\cite{Boehnke10} A first example for the dynamical, i.e.,
finite-frequency, case will be discussed in Sec.~\ref{sec:sw}.

\subsection{Antiferromagnetic susceptibility of the three-dimensional Hubbard model\label{sec:afm3d}}

In order to benchmark our approach, we investigate the antiferromagnetic
susceptibility of the half-filled Hubbard model \eqref{eqn:Hubbard} on a cubic lattice within the DMFT framework. All quantities are again expressed in units of the hopping $t$ and with $U/t=20$ and $T/t = 0.45$.
This temperature is sufficiently close to the DMFT N\'eel temperature
$T_\text{N}\approx 0.30t$ to yield a dominant vertex part, while still having a non-negligible bubble contribution.

%~~~~~~~~~~~~~~~~~~~~~~~~~~~~~~~~~~~~~~~~~~~~~~~~~~~~~~~~~~~~~~~~~~~~~~~~~~~~~~~
\begin{figure}[!Htb]
\begin{center}
  \noindent\includegraphics[scale=.65,angle=0]{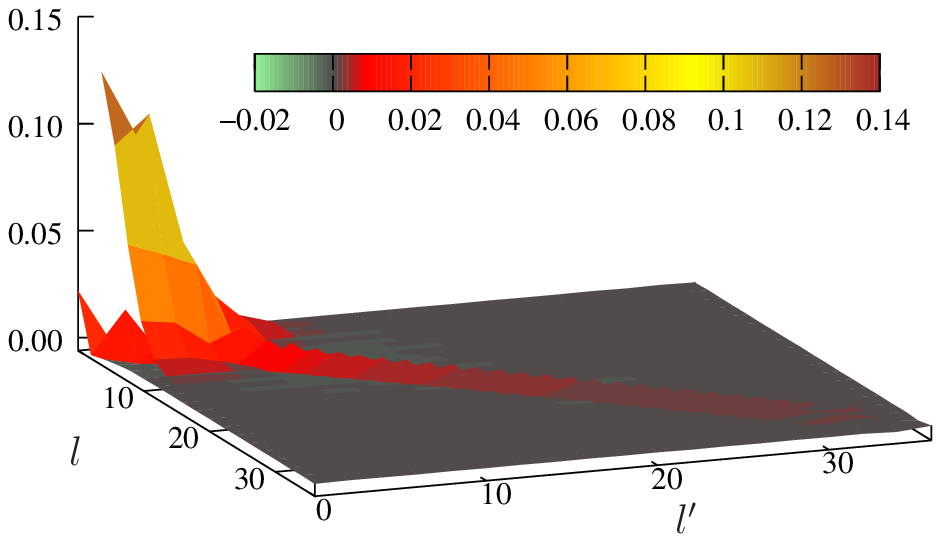}
  \noindent\includegraphics[scale=.65,angle=0]{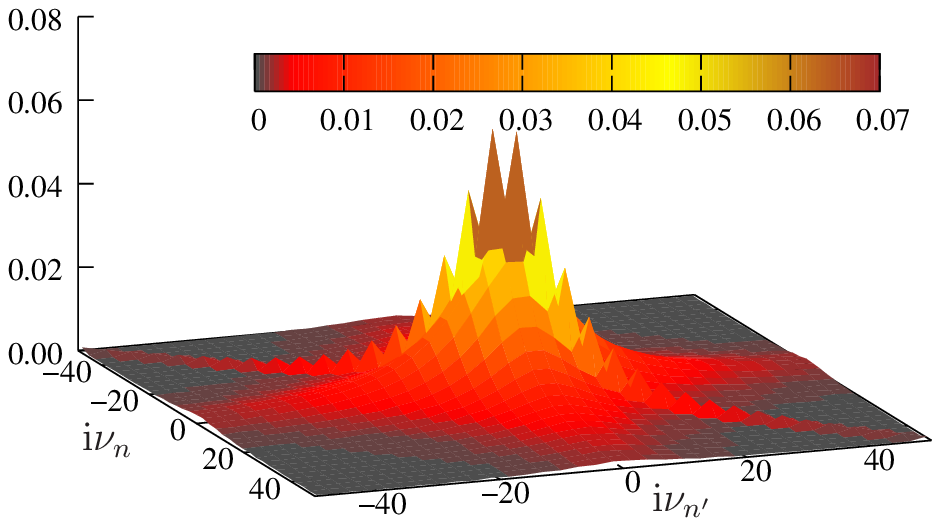}
\end{center}
\caption{(Color online) Generalized local magnetic susceptibility $\widetilde{\chi}^\mathrm{m}_\mathrm{loc}=\frac{1}{2}(\widetilde{\chi}^{\up\up}_\mathrm{loc}-\widetilde{\chi}^{\up\down}_\mathrm{loc})$ at the bosonic frequency $\i\omega_m=0$
         computed from the DMFT impurity problem.
         Upper panel: coefficients $\widetilde{\chi}^\mathrm{m}_{\mathrm{loc},ll^\prime}(0)$ in the mixed Legendre-Fourier
         representation. Lower panel: Fourier coefficients $\widetilde{\chi}^\mathrm{m}_\mathrm{loc} (\i\nu_n,\i\nu_{n^\prime},0)$.}
\label{fig:chilocal}
\end{figure}
%~~~~~~~~~~~~~~~~~~~~~~~~~~~~~~~~~~~~~~~~~~~~~~~~~~~~~~~~~~~~~~~~~~~~~~~~~~~~~~~

We compute the susceptibility $\widetilde{\chi}_\mathrm{loc}$ of the DMFT impurity problem
using the CT-HYB algorithm. In Fig.~\ref{fig:chilocal}
we compare the mixed Legendre-Fourier coefficients $\widetilde{\chi}_{ll^\prime}(\i\omega_m)$
to the Fourier coefficients $\widetilde{\chi} (\i\nu_n,\i\nu_{n^\prime},\i\omega_m)$. For clarity, we
focus on the first bosonic frequency $\i\omega_m=0$. We observe that the
$\widetilde{\chi}_{ll^\prime}(0)$ have a very fast decay except in the $l=l^\prime$
direction. This contrasts with the behavior of $\widetilde{\chi}
(\i\nu_n,\i\nu_{n^\prime},0)$ which exhibits slower decay in the three major directions $\i\nu_n=0$,
$\i\nu_{n^\prime}=0$ and $\i\nu_n=\i\nu_{n^\prime}$.

The generalized susceptibility in $\tau$-differences \eqref{eqn:chitau} has discontinuities along the planes $\tau_{14}=0$ and $\tau_{14}=\tau_{12}+\tau_{34}$ as well as non-analyticities (kinks) for $\tau_{12}=0$ and $\tau_{34}=0$. These planes induce corresponding slow decay in the Fourier representation \eqref{eqn:chifouriervschilegendre}.\cite{gunnarsson} When it comes to the mixed Legendre-Fourier representation \eqref{eqn:chitaullw2} however, the planes $\tau_{12}=0$ and $\tau_{34}=0$ are on the border of the imaginary-time region being expanded in this basis, which renders the coefficients insensitive toward these.

Computing lattice susceptibilities from Eq.~\eqref{eqn:chiq}, it is necessarily required to truncate the matrices.
This leads to difficulties when computing the susceptibility from the Fourier coefficients $\widetilde{\chi}_\mathrm{loc}(\i\nu_n,\i\nu_{n^\prime},\i\omega_m)$.
As we can see from Fig.~\ref{fig:chilocal}, the Fourier coefficients have a slow decay along three directions. The inversion of
$\widetilde{\chi}_\mathrm{loc}(\i\nu_n,\i\nu_{n^\prime},\i\omega_m)$ is delicate because many
coefficients are involved even for large $\nu,\nu^\prime$. One needs to use a very
large cutoff to obtain a precise result. Alternatively, one can try to separate the high- and low-frequency parts of the equation and replace the susceptibilities with their asymptotic form at high frequency (see Ref.~\onlinecite{kunes}). While is is effectively possible to treat larger matrices, it is still required to impose a cutoff on the high-frequency part for the numerical computations.

In the mixed Legendre-Fourier representation, the situation is different. Only
the coefficients along the diagonal decay slowly. In the inversion of
the matrix, the elements on the diagonal for large $l$ are essentially
recomputed from themselves. One can expect that there will be a lot less mixing
and thus a much faster convergence as a function of the truncation.

%~~~~~~~~~~~~~~~~~~~~~~~~~~~~~~~~~~~~~~~~~~~~~~~~~~~~~~~~~~~~~~~~~~~~~~~~~~~~~~~
\begin{figure}[!Htb]
\begin{center}
  \noindent\includegraphics[scale=.65,angle=0]{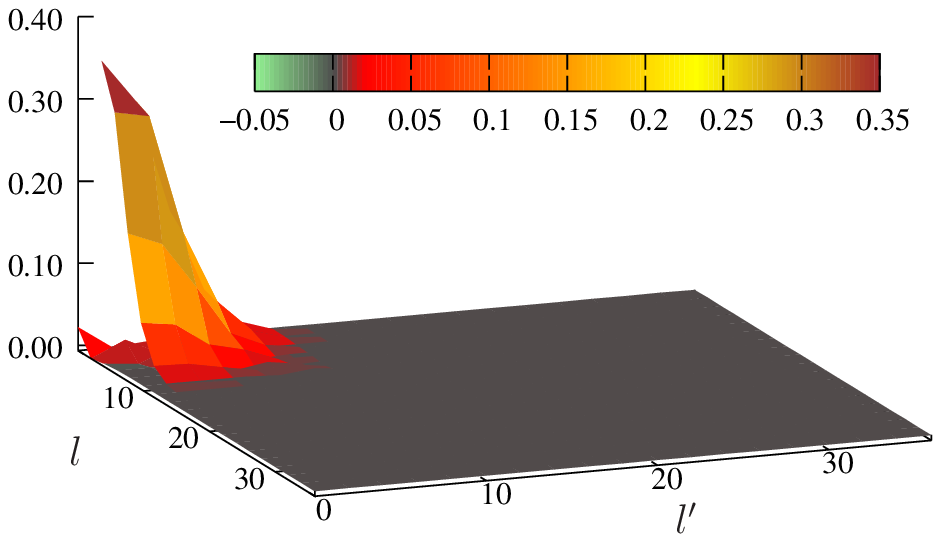}
  \noindent\includegraphics[scale=.65,angle=0]{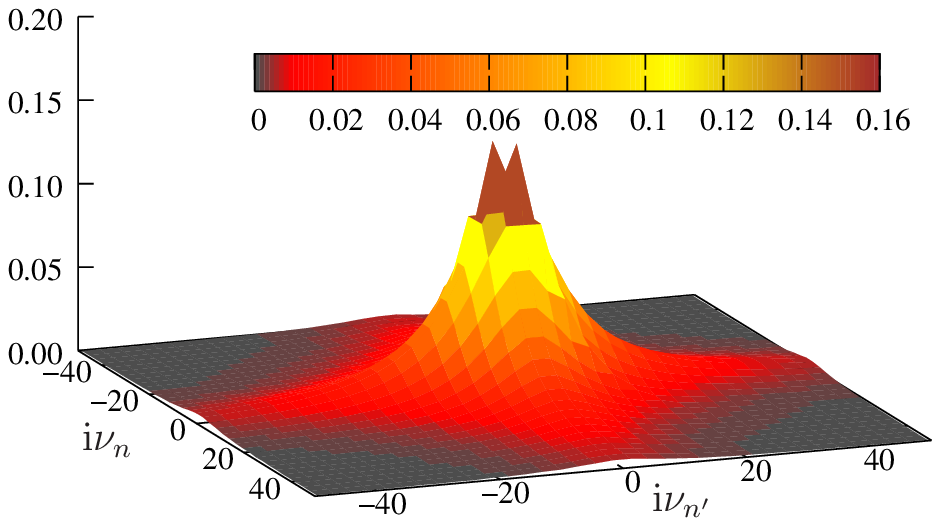}
\end{center}
\caption{(Color online) Vertex part of the generalized magnetic lattice susceptibility $\widetilde{\chi}^\mathrm{m}_\mathrm{latt} - \widetilde{\chi}^0_\mathrm{latt}$
         at the bosonic frequency $\i\omega_m=0$ and at the antiferromagnetic wave vector $\bq=(\pi,\pi,\pi)$.
         Upper panel: coefficients $\widetilde{\chi}^\mathrm{m}_{\mathrm{latt},ll^\prime}(0)-\widetilde{\chi}^0_{\mathrm{latt},ll^\prime}(0)$ in the mixed Legendre-Fourier
         representation. Lower panel: Fourier coefficients $\widetilde{\chi}^\mathrm{m}_\mathrm{latt} (\i\nu_n,\i\nu_{n^\prime},0)-\widetilde{\chi}^0_\mathrm{latt} (\i\nu_n,\i\nu_{n^\prime},0)$
         of the lattice susceptibility. Both plots employ the same number of coefficients.}
\label{fig:chilattice}
\end{figure}
%~~~~~~~~~~~~~~~~~~~~~~~~~~~~~~~~~~~~~~~~~~~~~~~~~~~~~~~~~~~~~~~~~~~~~~~~~~~~~~~

In Fig.~\ref{fig:chilattice}, we display the vertex part of the generalized lattice
susceptibility $\widetilde{\chi}_\mathrm{latt} -
\widetilde{\chi}^0_\mathrm{latt}$ obtained from Eq.~\eqref{eqn:chiq} in both
representations. In both cases, we see that the diagonal part quickly becomes
very small. In other words, the diagonal of the lattice susceptibility is
essentially given by the bubble part
$\widetilde{\chi}^0_\mathrm{latt}$. However, while essentially all the
information is condensed close to $l,l^\prime=0$ in the mixed Legendre-Fourier
representation, the Fourier coefficients still have a slow decay along the
directions given by $\i\nu_n=0$
and $\i\nu_{n^\prime}=0$. From this figure one can speculate that a quantity computed from
the Legendre-Fourier coefficients will converge rapidly as a  function
of a cutoff $l_\mathrm{max}$. However, we need to make sure that the
coefficients close to $l,l^\prime=0$ are not affected much by the truncation.

%~~~~~~~~~~~~~~~~~~~~~~~~~~~~~~~~~~~~~~~~~~~~~~~~~~~~~~~~~~~~~~~~~~~~~~~~~~~~~~~
\begin{figure}[!Htb]
\begin{center}
  \includegraphics[scale=.65,angle=0]{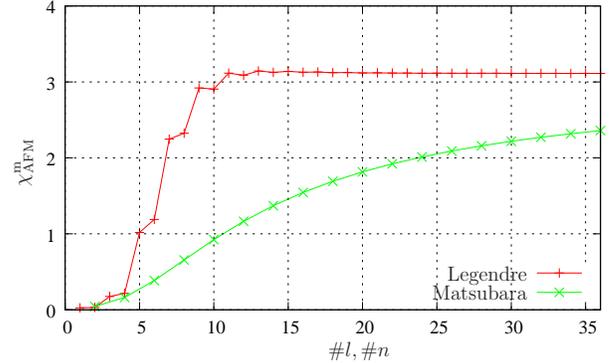}
\end{center}

\caption{(Color online) Antiferromagnetic susceptibility as a function of the number of Legendre ($\#l=l_{\text{max}}+1$) and Matsubara ($\#n=2n_{\text{max}}+2$) coefficients, respectively, used in the calculation.}
\label{fig:chistatic}
\end{figure}
%~~~~~~~~~~~~~~~~~~~~~~~~~~~~~~~~~~~~~~~~~~~~~~~~~~~~~~~~~~~~~~~~~~~~~~~~~~~~~~~

In order to assess the validity of these speculations we compute the static antiferromagnetic ($\bq=(\pi,\pi,\pi)$)
susceptibility $\chi^\mathrm{m}(0,\bq)$ as a function of the cutoff in both representations. It is obtained from Eq.~\eqref{eqn:chisuml}
using the magnetic susceptibility $\widetilde{\chi}^\mathrm{m}=\frac{1}{2}(\widetilde{\chi}^{\up\up}-\widetilde{\chi}^{\up\down})$.

Since the diagonal of the lattice susceptibility is essentially given by the
bubble (see Fig.~\ref{fig:chilattice}), the sums above are
performed in two steps. The vertex part shown in Fig.~\ref{fig:chilattice} is
summed up to the chosen cutoff, while the bubble part is summed over
all frequencies with the knowledge of its high-frequency behavior.  The result
is shown in Fig.~\ref{fig:chistatic}. It reveals a major benefit of the
Legendre representation: the susceptibility converges much faster as a function of the cutoff. The static susceptibility is essentially converged at $l_\mathrm{max}
\sim 12$. This corroborates the idea that the small-$l,l^\prime$ part of
$\widetilde{\chi}_{\mathrm{latt},ll^\prime}$ is only weakly dependent on the further
diagonal elements of $\widetilde{\chi}_{\mathrm{loc},ll^\prime}$.

\subsection{Dynamical susceptibility of the two-dimensional Hubbard model}
\label{sec:sw}

\begin{figure}[!Htb]
\begin{center}
  \includegraphics[scale=.65,angle=0]{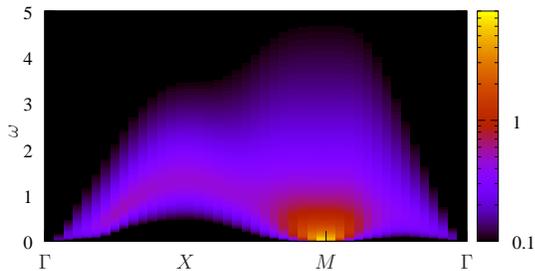}
\end{center}
\caption{(Color online) Imaginary part of the magnetic susceptibility on the real frequency axis along high-symmetry lines in the 2D Brillouin zone.}
\label{fig:sw}
\end{figure}

As a final benchmark, we demonstrate that our method is not restricted to the
static case. To this end, we show the momentum resolved dynamical magnetic
susceptibility $\chi(\omega,\bq)$ for a DMFT calculation for the half-filled two-dimensional (2D)
square lattice Hubbard model in Fig. \ref{fig:sw}.  We have chosen an on-site
interaction $U/t=4$ and temperature $T/t=0.25$, which is slightly above the
DMFT N\'eel temperature.  The susceptibility was computed from the Legendre
representation according to Eq.~\eqref{eqn:chisuml} using $20 \times 20$
Legendre coefficients, which was sufficient for all bosonic frequencies. In
general, for higher bosonic frequencies more Legendre coefficients are needed
to represent the vertex part of the generalized magnetic lattice
susceptibility. However, no additional structure appears in the high $l$, $l^\prime$
region. We then analytically continued the data using Pad\'e
approximants.~\cite{pade} The figure shows the typical magnon
spectrum~\cite{Preuss97,Hochkeppel08,ldfa} reminiscent of a spin wave in this
paramagnetic state with strongly enhanced weight at the antiferromagnetic wave
vector $\bq=(\pi,\pi)$ due to the proximity of the mean-field antiferromagnetic
instability.

%~~~~~~~~~~~~~~~~~~~~~~~~~~~~~~~~~~~~~~~~~~~~~~~~~~~~~~~~~~~~~~~~~~~~~~~~~~~~~~~
%~~~~~~~~~~~~~~~~~~~~~~~~~~~~~~~~~~~~~~~~~~~~~~~~~~~~~~~~~~~~~~~~~~~~~~~~~~~~~~~

\section{Conclusion}

In this paper, we have studied the representation of imaginary-time Green's functions in terms of a Legendre orthogonal polynomial basis.
We have shown that CTQMC can directly accumulate the Green's function in this basis. This representation has several advantages over the standard Matsubara frequency representation: (i) It is much more compact, i.e., coefficients decay much faster; this is particularly interesting for storing and manipulating the two-particle Green's functions. Moreover,  two-particle response functions can be computed directly in the Legendre representation, without the need to transform back to the Matsubara representation. 
In particular, the matrix manipulations required for the solution of the Bethe-Salpeter equations can be performed in this basis. We have shown that this greatly enhances the accuracy of the calculations, since in contrast to the Matsubara representation the error due to the truncation of the matrices becomes negligible. 
(ii) The Monte Carlo noise is mainly concentrated in the higher Legendre coefficients, the contribution of which is usually very small; this allows us to develop a systematic method to filter out noise in physical quantities and to obtain more accurate values for, e.g., the correlation  energy in LDA+DMFT computations.

\acknowledgments
L.B. and F.L. thank A.I.~Lichtenstein for helpful discussions. M.F. and O.P. thank M.~Aichhorn for helpful discussions and providing the parameters and data
from Ref.~\onlinecite{aichhorn2009}.  O.P. thanks J.M.~Normand for pointing out
Ref.~\onlinecite{BatemanBook}. We thank P.~Werner for a careful reading of the manuscript.  Calculations were performed with the TRIQS\cite{TRIQS} hier project using HPC resources
from The North-German Supercomputing Alliance (HLRN) and from GENCI-CCRT (Grant
No. 2011-t2011056112). TRIQS uses some libraries of the ALPS~\cite{ALPS20} project. This work was supported by the DFG Research Unit FOR
1346.

\begin{appendix}

\section{SOME PROPERTIES OF THE LEGENDRE POLYNOMIALS}
\label{sec:propertieslegendre}
In this appendix, we summarize for convenience some basic properties of the Legendre polynomials. Further references can be found in 
Refs.~\onlinecite{GradshteynBook,WhittakerWatsonBook,BatemanBook}.
We use the standardized polynomial $P_l(x)$ defined on $x\in [-1,1]$ 
through the recursive relation
\begin{align}
  (l+1)P_{l+1}(x)&=(2l+1)xP_l(x)-lP_{l-1}(x) \\
P_0(x)&=1,\quad P_1(x) = x
  \label{App.def.LegendrePoly}
\end{align}
$P_l$ are orthogonal and their normalization is given by
\begin{equation}
  \int_{-1}^1 \d x P_k(x) P_l(x) = \frac{2}{2l+1} \delta_{kl}
  \label{App.A.LegendreOrthogonality}
\end{equation}
The $P_l$ are bounded on the segment $[-1,1]$ by~\cite{WhittakerWatsonBook}
\begin{equation}
  |P_l(x)|\leq1 
  \label{App.A.PlBounded}
\end{equation}
with the special points 
\begin{equation}
  P_l(\pm1)=(\pm1)^l.
  \label{App.A.PlBorders}
\end{equation}
The primitive of $P_l(x)$ that vanishes at $x=-1$ is
(cf. Ref.~\onlinecite{BatemanBook}, Vol II, section 10.10)
\begin{equation}
  \int_{-1}^x \d yP_l(y)=\frac{P_{l+1}(x)-P_{l-1}(x)}{2l+1}, \qquad l\geq 1
\label{App.A.PlPrimitive}
\end{equation}
By orthogonality or \eqref{App.A.PlBorders}, it also vanishes at $x=1$.
The Fourier transform of the Legendre polynomial restricted to the segment $[-1,1]$ is given by
formula 7.243.5 of Ref.~\onlinecite{GradshteynBook}
\begin{align}
  \int_{-1}^1 e^{i a x } P_l(x) \, dx &= i^l \sqrt{\frac{2 \pi}{a}} J_{l+ \frac{1}{2}} (a)\nonumber\\
&=2i^lj_l(a),
  \label{FourierTransformLegendreAsBessel}
\end{align}
where $J$ denotes the Bessel function and $j_l(a)=\sqrt{\frac{\pi}{2a}}J_{l+\frac{1}{2}}(a)$ denotes the spherical Bessel functions.

\section{FAST DECAY OF THE LEGENDRE COEFFICIENTS}
\label{sec:Rapiddecayproof}
Let us consider a function $g(\tau)$ smooth on the segment $[0,\beta]$ (i.e. to be precise ${\cal C}^\infty$, indefinitely differentiable), and $\beta-$antiperiodic, like a Green's function.
In this appendix, we show that its Legendre coefficients decay faster than any
power law contrary to its standard Fourier expansion coefficients which decay
as power laws determined by the discontinuities of the function and its derivatives.

Let us start by reminding the asymptotics of the standard Fourier expansion coefficients 
on fermionic Matsubara frequencies.
These coefficients are given by
\begin{align}
  \hat g(\i \nu_n) &= \int_0^\beta d \tau\ g(\tau) e^{\i \nu_n \tau} \\
  &= \frac{ \left. g(\tau) e^{\i \nu_n \tau} \right |_0^\beta}{\i\nu_n}  -  \int_0^\beta d \tau\ g'(\tau) \frac{e^{\i \nu_n \tau}}{\i\nu_n} 
  \label{}
\end{align}
The coefficients vanish for $n\rightarrow\infty$, and applying the same result to $g'$, one obtains
\begin{equation}\label{App.D.g_FourierAsymp}
  \hat g(\i \nu_n) = - \frac{ g(\beta^-) + g(0^+)}{\i\nu_n} + O\left(\frac{1}{\nu_n^2}\right)
\end{equation}
Let us now turn to the Legendre expansion. Using the same rescaling as before, we can consider for simplicity
a function $f(x)$ smooth on $[-1,1]$. We can proceed in a similar way using
the primitive of the Legendre polynomial (which is also given by a simple formula,  \eqref{App.A.PlPrimitive}).
For $l\geq 1$, we have 
\begin{align}
  \frac{f_l}{\sqrt{2l+1}} =& \int_{-1}^1 d x\ f(x) P_l(x) \nonumber\\
  =& \left. f(x)   \left(\int_{-1}^x \d y\ P_l(y) \right) \right |_{-1}^1 -
    \nonumber\\ &
    \int_{-1}^1 d x\ f'(x) 
   \left(\int_{-1}^x \d y\ P_l(y) \right) 
    \nonumber\\ =&-
    \int_{-1}^1 d x\ f'(x) 
    \frac{P_{l+1}(x) - P_{l-1}(x)}{2l+1} 
\end{align}
The crucial difference with the Fourier case is that, for $l\geq 1$, 
{\sl the boundary terms always cancel}, whatever the function $f$ due to the orthogonality property of the polynomials
(it can also be checked directly from  \eqref{App.A.PlBorders}).
So we are left with just the integral term.
Since the Legendre coefficients of $f'$ vanish at large $l$ (by applying the previous formula to $f'$), we get
instead of (\ref{App.D.g_FourierAsymp}) 
\begin{equation}
  \frac{f_l}{\sqrt{2l+1}} =  o\left(\frac{1}{l}\right)
\end{equation}
In both cases, the reasoning can be reproduced recursively, by further differentiating the function, as long as no singularity are encountered.
In the Fourier case, it produces the well-known high-frequency expansion in terms of the 
discontinuity of the function and its derivatives.
In the Legendre case, we find that the coefficients are $o(1/l^k)$ as soon as $f$ is $k$ times differentiable.
Hence if the function is smooth on $[-1,1]$, the coefficients decays asymptotically faster than any power law.

The only point that remains to be checked is that indeed $G(\tau)$ is smooth on $[0,\beta]$.
It is clear from its spectral representation 
\begin{equation}
G(\tau)= - \int_{-\infty}^\infty\d\nu\frac{\myexp{-\tau\nu}}{1+\myexp{-\beta\nu}}A(\nu)
\end{equation}
if we admit that the spectral function $A(\nu)$ has compact support, by differentiating under the integral.

Finally, while this simple result of ``fast decay'' is enough for our purposes in this paper, 
it is possible to get much more refined statements on the asymptotics of the Legendre coefficients
of the function $f$, in particular when it has some analyticity properties.
For a detailed discussion of these issues, and in particular of the conditions needed to get the 
generic exponential decay of the coefficients, we refer to Ref.~\onlinecite{BoydBook}.

\section{DIRECT ACCUMULATION OF THE LEGENDRE COEFFICIENTS FOR THE CT-HYB ALGORITHM}\label{App.GAccu}
\label{sec:accumulation}

In this appendix, we describe how to compute directly the Legendre expansion of the one-particle and the two-particle
Green's function.

\subsection{The accumulation formulas in CT-HYB}
For completeness, let us first recall the accumulation formula for the one-particle and the two-particle
Green's functions in the CT-HYB algorithm \cite{Werner06,Haule07, WernerPRBLong,ctqmcrmp}, which sums the perturbation theory in
the hybridization function $\Delta_{ab}(\i\nu_n)$ on the Matsubara axis.
While these formulas have appeared previously in the literature, this simple functional derivation
emphasizes the ``Wick''-like form of the high-order correlation function.

The partition function of the impurity model reads
\begin{equation}\label{eqn:app:Z}
Z = \int {\cal D} c^{\dagger} {\cal D} c \exp ( - S_{\text{eff}} )
\end{equation}
where the effective action has the form
\begin{gather}
 S_{\text{eff}} = - \iint_{0}^{\beta } \d \tau  \d \tau^\prime
 \sum_{\substack{A,B}}c^{\dagger }_{A} (\tau )
G_{0, AB}^{-1} (\tau ,\tau^\prime)
c^{ }_{B} (\tau^\prime )\nonumber
\\  + \int_{0}^{\beta } \d \tau \label{Seff}
H_\text{int}(\{c^\dagger_A(\tau),c_A(\tau)\})
\\
 G^{-1}_{0AB}(\i \nu_n) = (\i \nu_n + \mu) \delta_{AB} - h^0_{AB} - \Delta_{AB}(\i \nu_{n}),\label{eqn:app:G0}
\end{gather}
To simplify the notations, we use here a generic index $A,B$. In the case where there are symmetries,
like the spin $SU(2)$ symmetry in the standard DMFT problem, the Green's functions are block diagonal.
For example, the generic index $A$ can be $(a,\sigma)$, where $a$ is an orbital or site index,
and spin index $\sigma= \uparrow,\downarrow$ is the block index.

The partition function  is expanded in powers of the hybridization $\Delta$ as
\def\CC{ \prod_{i=1}^{n} c^{\dagger}_{\lambda_{i}}(\tau_{i}) c_{\lambda'_{i}}(\tau^\prime_{i})}
\begin{gather}
Z = \sum_{n\geq 0} 
\int \prod_{i=1}^{n} \d\tau_{i} \d \tau^\prime_{i}
\sum_{\lambda_{i}, \lambda'_i } w(n,\{ \lambda_j,\lambda'_j,\tau_j,\tau^\prime_j\})
\\
w(n,\{ \lambda_j,\lambda'_j,\tau_j,\tau^\prime_j\})  \equiv \frac{1}{n!^2}
\det_{1\leq i,j \leq n}
\bigl [
  \Delta_{\lambda_{i}, \lambda'_{j}}(\tau_{i} - \tau^\prime_{j})
\bigr] \times \nonumber \\
\mathop{\text{Tr}}
\left( \Ttau
\myexp{-\beta H_\mathrm{loc}} \CC
\right),
\end{gather}
where $\Ttau$ is time ordering and $H_{\mathrm{loc}}$ is the local Hamiltonian 
\cite{Werner06,Haule07, WernerPRBLong,ctqmcrmp}. 
$|w|$ are the weights of the Quantum Monte Carlo 
Markov chain.
Introducing the short notation $ {\cal C} \equiv (n, \{ \lambda_j,\lambda'_j,\tau_j,\tau^\prime_j\})$ for the QMC configuration, 
the partition function $Z$ and the average of any function $f$ over the configuration space (denoted by angular bracket in this section)
are given by 
\begin{align}
  Z &= \sum_{\cal C} w({\cal C}) \\ 
  \moy{ f({\cal C}) }
  &= \frac{1}{Z}  \sum_{\cal C} w({\cal C}) f({\cal C})
\end{align}
The one-particle and two-particle Green's functions are obtained as functional derivatives of $Z$ with 
respect to the hybridization function, as
\begin{subequations}
  \label{App.B.G_as_Z_deriv}
\begin{align}
  G_{AB}(\tau_1,\tau_2) &= - \frac{1}{Z} \frac{\partial Z}{ \partial \Delta_{BA}(\tau_2,\tau_1)}
  \\
  G^{(4)}_{ABCD}(\tau_1,\tau_2, \tau_3, \tau_4) &= \frac{1}{Z} \frac{\partial^2 Z}{ \partial \Delta_{BA}(\tau_2,\tau_1)\partial \Delta_{DC}(\tau_4,\tau_3)}
\end{align}
\end{subequations}
\def\genMat{\bar\Delta}
\def\DeltaInv{\bar M}
In order to use the expansion of $Z$, we need to compute the derivative of a determinant with respect 
to its elements. Let us consider a general matrix $\genMat$, its inverse $\DeltaInv \equiv \genMat^{-1}$
and use Grassman integral representation
\begin{equation}
 \det
 \genMat
 =  \int \prod_{i}d\eta_i\d\bar\eta_i \myexp{\sum_{ij}\bar\eta_i \genMat_{ij}\eta_j}
  \label{App.B.detRepre}
\end{equation}
Using the Wick theorem, we have
\begin{subequations}
  \label{App.B.DerivationOfDet}
\begin{align}
\nonumber
\frac{\partial \det  \genMat}{\partial \genMat_{ba} } &= 
  \int \prod_{i}\d \eta_i\d \bar\eta_i \bigl( \bar\eta_b\eta_a\bigr) \myexp{\sum_{ij}\bar\eta_i\bar \Delta_{ij}\eta_j}
\\
&= \det \bar\Delta \times \DeltaInv_{ab}
\\
\nonumber
\frac{\partial^2 \det \bar\Delta}{\partial \bar\Delta_{ba}\partial \bar\Delta_{dc} } &= 
  \int \prod_{i}\d \eta_i\d \bar \eta_i \bigl(\bar\eta_b\eta_a \bar\eta_d\eta_c \bigr) \myexp{\sum_{ij}\bar\eta_i\bar\Delta_{ij}\eta_j}
\\
&= \det \bar\Delta \left( \DeltaInv_{ab}\DeltaInv_{cd} - \DeltaInv_{ad}\DeltaInv_{cb} \right)
\end{align}
\end{subequations}
Let us now apply (\ref{App.B.DerivationOfDet}) by introducing for each configuration
 $ {\cal C} \equiv (n, \{ \lambda_j,\lambda'_j,\tau_j,\tau^\prime_j\})$ 
 the matrix $\hat \Delta({\cal C})$ of size $n$  given by 
\begin{equation}
  \hat \Delta({\cal C})_{ij} \equiv \Delta_{\lambda_{i}, \lambda'_{j}}(\tau_i - \tau'_j)
  \label{AppB_Def_HatDelta}
\end{equation}
and its inverse $M^{\cal C}\equiv \bigl(\hat\Delta({\cal C}) \bigr)^{-1}$.
We obtain
\begin{subequations}
\begin{align}\!\!\!
  \frac{\partial w({\cal C})}{\partial \Delta_{BA}(\tau_2,\tau_1) }
  &= \frac{ w({\cal C}) }{\det \hat \Delta({\cal C}) }  \sum_{\alpha,\beta=1}^n \frac{\partial \det \hat \Delta({\cal C})}{ \partial  \hat \Delta({\cal C})_{\beta\alpha} } 
    \frac{\partial \hat \Delta({\cal C})_{\beta\alpha} }{\partial \Delta_{BA}(\tau_2,\tau_1)} 
\nonumber\\
  &=  w({\cal C}) \sum_{\alpha,\beta=1}^n  M^{\cal C}_{\alpha\beta}
    \frac{\partial \hat \Delta({\cal C})_{\beta\alpha} }{\partial \Delta_{BA}(\tau_2,\tau_1)}   
  \end{align}
  and 
\begin{multline}
  \frac{\partial^2 w({\cal C})}{\partial \Delta_{BA}(\tau_2,\tau_1) \partial \Delta_{DC}(\tau_4,\tau_3) }
  = \frac{ w({\cal C}) }{\det \hat \Delta({\cal C}) }
 \times \\
  \sum_{\alpha\beta\gamma\delta=1}^n \frac{\partial^2 \det \hat \Delta({\cal C})}{ \partial  \hat \Delta({\cal C})_{\beta\alpha} 
   \partial  \hat \Delta({\cal C})_{\delta\gamma} } 
 \frac{\partial \hat \Delta({\cal C})_{\beta\alpha} }{\partial \Delta_{BA}(\tau_2,\tau_1)} 
    \frac{\partial \hat \Delta({\cal C})_{\delta\gamma} }{\partial \Delta_{DC}(\tau_4,\tau_3)}
  \end{multline}
\end{subequations}
Denoting
\begin{align}
  D({\cal C})_{AB\tau_1\tau_2}^{\alpha\beta} &\equiv 
 \frac{\partial \hat \Delta({\cal C})_{\beta\alpha} }{\partial \Delta_{BA}(\tau_2,\tau_1)} \nonumber
 \\ &=
   \delta(\tau_1-\tau'_\alpha)\delta(\tau_2-\tau_\beta) \delta_{\lambda'_\alpha,A} \delta_{\lambda_\beta,B}
\end{align}
we finally obtain the accumulation formulas for the Green's functions \cite{Werner06, WernerPRBLong}
\begin{subequations}
\begin{gather}
  \label{App.B.G_accu}
 G_{AB}(\tau_1,\tau_2)= -
 \left<\sum_{\alpha\beta=1}^n 
 M_{\alpha\beta}^{{\cal C}}
 D({\cal C})_{AB\tau_1\tau_2}^{\alpha\beta} 
\right>
 \\
 \nonumber 
 G^{(4)}_{ABCD}(\tau_1,\tau_2,\tau_3,\tau_4)=
 \biggl<\sum_{\alpha\beta\gamma\delta=1}^n
(M^{{\cal C}}_{\alpha\beta}M^{{\cal C}}_{\gamma\delta}-M^{{\cal C}}_{\alpha\delta}M^{{\cal C}}_{\gamma\beta}) \times  \\ 
  \label{App.B.G_accu4}
 D({\cal C})_{AB\tau_1\tau_2}^{\alpha\beta} 
 D({\cal C})_{CD\tau_3\tau_4}^{\gamma\delta}  
\biggr>
\end{gather}
\end{subequations}

\subsection{Legendre expansion of the one particle Green's function}
We take into account the  time translation invariance and 
the $\tau$-antiperiodicity of the Green's function in the following way.
A priori, in \eqref{App.B.G_accu}, the arguments $\tau_1, \tau_2$ are in the interval $[0,\beta]$.
We can however easily make this function $\beta-$antiperiodic in both arguments
\begin{multline}
\widetilde{G}_{AB}(\tau_1,\tau_2)=\\ -
 \left<\sum_{\alpha\beta=1}^n 
 M_{\alpha\beta}^{{\cal C}}
 \delta^-(\tau_1-\tau'_\alpha) \delta^-(\tau_2-\tau_\beta)
\delta_{\lambda'_\alpha,A} \delta_{\lambda_\beta,B}
\right>
\end{multline}
where we defined the periodic and antiperiodic Dirac comb respectively by
\begin{equation}
\label{def_anti_Dirac_comb}
  \delta^\pm(\tau) \equiv \sum_{n\in \mathbb{Z}} (\pm1)^n \delta(\tau - n\beta)
\end{equation}

At convergence of the Monte-Carlo Markov chain, the Green's function is in fact translationally invariant in imaginary time
and we have 
\begin{equation}
  G_{AB}(\tau) = \frac{1}{\beta} \int_0^\beta d s\ \widetilde{G}_{AB}(\tau + s,s)
\end{equation}
which leads to 
\begin{equation}\label{eqn:accuG2}
  G_{AB}(\tau)= - \frac{1}{\beta}
 \left<\sum_{\alpha\beta=1}^n 
 M_{\alpha\beta}^{{\cal C}}
\delta^-\bigl (\tau - (\tau'_\alpha-\tau_\beta) \bigr)
\delta_{\lambda'_\alpha,A} \delta_{\lambda_\beta,B}
\right>
\end{equation}
Finally, Eq.~\eqref{eqn:accuG2} can be transformed to a measurement 
in the Legendre representation according to \eqref{eqn:G2leg}
\begin{equation}
  G_{AB;l}=-\frac{\sqrt{2l+1}}{\beta}
\left< 
\sum_{\alpha\beta=1}^n 
 M_{\alpha\beta}^{{\cal C}}
\widetilde{P}_l(\tau'_\alpha-\tau_\beta)
\delta_{\lambda'_\alpha,A} \delta_{\lambda_\beta,B}
\right>
\end{equation}
where $\widetilde{P}(\delta\tau)$ is defined by  
\begin{equation}\label{eqn:Ptilde}
  \widetilde{P}_l(\delta\tau)= 
\begin{cases}
 \phantom{-}P_l(x(\delta\tau))&\delta\tau>0\\
 -P_l(x(\delta\tau+\beta))&\delta\tau<0
\end{cases}
\end{equation}

\subsection{Legendre accumulation of the two-particle Green's function}

The generalized susceptibility $\widetilde{\chi}$ of \eqref{eqn:chitau}
can be expressed in term of $G$ and $G^{(4)}$   as 
\begin{align}
\widetilde\chi_{abcd}^{\sigma\sigma^\prime}(\tau_{12},\tau_{34},\tau_{14})=&
G^{(4)}_{b\sigma,a\sigma,d\sigma^\prime,c\sigma^\prime}(\tau_{21},\tau_{43},\tau_{23})\nonumber\\
&-G_{b\sigma,a\sigma}(\tau_{21})G_{d\sigma^\prime\!,c\sigma^\prime}(\tau_{43})
\end{align}
so in this subsection we will focus on the computation of $G^{(4)}$.
We take into account the  time translation invariance with the same technique as for the one-particle Green's function.
First we make the function $G^{(4)}(\tau_1,\tau_2,\tau_3,\tau_4)$ fully $\beta-$antiperiodic in the 
four variables using the antiperiodic Dirac comb $\delta^-$ defined in \eqref{def_anti_Dirac_comb}, and we use  
the time translation invariance of the Green's function to obtain 

  \begin{multline}
  G^{(4)}(\tau_{12},\tau_{34},\tau_{14})\\= \frac{1}{\beta}
\int_0^\beta {d\bar \tau} \ 
  \widetilde{G}^{(4)}(\tau_{14} + \bar \tau,
  \tau_{14} - \tau_{12} + \bar \tau  ,
  \tau_{34} + \bar \tau,\bar \tau)
\end{multline}
From (\ref{App.B.G_accu4}), we get 
\begin{widetext}
\begin{multline} 
G^{(4)}_{ABCD}\!(\tau_{12}, \tau_{34}, \tau_{14}) =  
\frac{1}{\beta}
 \biggl<\sum_{\alpha\beta\gamma\delta=1}^n
(M^{{\cal C}}_{\alpha\beta}M^{{\cal C}}_{\gamma\delta}-M^{{\cal C}}_{\alpha\delta}M^{{\cal C}}_{\gamma\beta})
\times  \\ 
\delta^-\bigl(\tau_{12} -( \tau'_\alpha - \tau_\beta) \bigl)
\, \delta^-\bigl(\tau_{34} - ( \tau'_\gamma - \tau_\delta) \bigl)
\, \delta^+\bigl(\tau_{14} -( \tau'_\alpha - \tau_\delta) \bigl)
\  \delta_{\lambda'_\alpha,A} \delta_{\lambda_\beta,B} \ \delta_{\lambda'_\gamma,C} \delta_{\lambda_\delta,D}
\Bigg>
\end{multline} 
where $\delta^+$ and $\delta^-$  are defined in \eqref{def_anti_Dirac_comb}.
Applying~\eqref{eqn:chitaullw2}, the accumulation formula in the mixed Legendre-Fourier basis is straightforwardly obtained as 
\begin{multline}
  G^{(4)}_{ABCD}(l,l',\i\omega_m)=\frac{\sqrt{2l+1}\sqrt{2l^\prime+1}}{\beta}(-1)^{l^\prime+1}
\times  \\
  \biggl<\sum_{\alpha\beta\gamma\delta=1}^n
(M^{{\cal C}}_{\alpha\beta}M^{{\cal C}}_{\gamma\delta}-M^{{\cal C}}_{\alpha\delta}M^{{\cal C}}_{\gamma\beta}) 
\widetilde{P}_l\left(\tau'_\alpha - \tau_\beta\right)\widetilde{P}_{l^\prime}\left(\tau'_\gamma - \tau_\delta
\right)\myexp{\i\omega_m ( \tau'_\alpha - \tau_\delta) }
 \delta_{\lambda'_\alpha,A} \delta_{\lambda_\beta,B}\delta_{\lambda'_\gamma,C} \delta_{\lambda_\delta,D}
\Bigg>
\label{eqn:app:legendre2pmeas}
\end{multline}
where $\widetilde{P}$ is defined in \eqref{eqn:Ptilde}.
\end{widetext}

We note that the measurement can be factorized to speed up the measurement
process. In the Legendre measurement, only the part involving the first product of $M$-matrices factorizes, as can be seen from \eqref{eqn:app:legendre2pmeas}.
Note, however, that the second product of $M$-matrices merely generates crossing symmetry, so that the full information on this quantity is already contained in the first term. Hence this symmetry can be reconstructed after the simulation.
In the one band case, the second product is proportional to $\delta_{\sigma\sigma^\prime}$, so that the $G^{(4)\uparrow\downarrow}$-component can be measured directly.
For the $G^{(4)\uparrow\uparrow}$-component we only measure the term proportional to $M^{\mathcal{C}}_{\uparrow}M^{\mathcal{C}}_{\uparrow}$ and construct this component by antisymmetrization afterwards.

\section{ACCUMULATION FORMULA FOR THE CT-INT AND CT-AUX ALGORITHMS}
\label{app:CTINTacc}
Using a notation in analogy to the previous section, the expansion of the partition function $Z$, Eqs. (\ref{eqn:app:Z}-\ref{eqn:app:G0}), in the continuous-time interaction expansion (CT-INT) method\cite{Rubtsov05} is given by
\begin{gather}
Z = \sum_{n\geq 0} \int \prod_{i=1}^{n} \d\tau_{i}
\sum_{
\substack{
\lambda_{2i-1}, \lambda'_{2i-1}\\
\lambda_{2i}, \lambda'_{2i}}
}
w(n,\{ \lambda_j,\lambda'_j,\tau_j\})
\\
w(n,\{ \lambda_j,\lambda'_j,\tau_j\}) \equiv \frac{1}{n!}
\det_{1\leq i,j \leq 2n}
\bigl [
G_{0\lambda_{i}, \lambda'_{j}}(\bar{\tau}_{i} - \bar{\tau}_{j})
\bigr] \times \nonumber \\
\times \prod_{i=1}^{n} U_{\lambda_{2i-1} \lambda'_{2i-1}\lambda_{2i} \lambda'_{2i} },
\end{gather}
where $\bar{\tau}_{i}\equiv \tau_{\lfloor(i+1)/ 2\rfloor}$ and we have assumed the interaction part of the Hamiltonian to be of the form
$
H_{\text{int}}(\{c^{\dagger}_{A},c_{A}\}) = \sum_{ABCD} U_{ABCD} c^{\dagger}_{A}c_{B}c^{\dagger}_{C}c_{D} $ and $A=(a,\sigma)$ is a generic index with $a$ being the orbital or site index and $\sigma=\uparrow,\downarrow$ the spin index.
In the CT-INT algorithm, we propose to measure the Legendre coefficients of $S\equiv\Sigma G$ based on the self-energy binning measurement originally introduced for the continuous-time auxiliary field (CT-AUX) algorithm\cite{CTAUX}. Introducing the matrix
\begin{align}
\hat{G}_{0}(\mathcal{C})_{ij} = G_{0\lambda_{i}\lambda'_{j}}(\bar{\tau}_{i}-\bar{\tau}_{j})
\end{align}
and its inverse, $M^\mathcal{C}\equiv(\hat{G}_{0}(\mathcal{C}))^{-1}$, the self-energy binning measurement for the CT-INT can be written as
\begin{align}
S_{AB}(\tau) = -\left\langle \sum_{\alpha\beta=1}^{2n} \delta(\tau-\bar{\tau}_{\alpha}) \delta_{A\lambda'_{\alpha}} M^{\mathcal{C}}_{\alpha\beta} G^{0}_{\lambda_{\beta}B}(\bar{\tau}_{\beta}) \right\rangle .
\end{align}
This can be straightforwardly transformed to a measurement in the Legendre basis by applying \eqref{eqn:G2leg}:
\begin{align}
S_{AB,l} = -\sqrt{2l+1}\left\langle \sum_{\alpha\beta=1}^{2n} \delta_{A\lambda'_{\alpha}}P_{l}(x(\bar{\tau}_{\alpha})) M^{\mathcal{C}}_{\alpha\beta} G^{0}_{\lambda_{\beta}B}(\bar{\tau}_{\beta}) \right\rangle .
\end{align}
An analogous formula also applies to the CT-AUX.
In practice, translational invariance may be used to generate multiple estimates for $S$ within a given configuration. The Green's function is obtained by transforming $S$ to Matsubara representation and using Dyson's equation. The moments of $G$ are straightforwardly computed from the moments of $\Sigma G$ and the knowledge of those of $G_0$.

\section{EXPLICIT FORMULA FOR  $\mathbf{T_{n l}}$ AND ITS HIGH FREQUENCY EXPANSION}
\label{sec:Tnulformula}

%~~~~~~~~~~~~~~~~~~~~~~~~~~~~~~~~~~~~~~~~~~~~~~~~~~~~~~~~~~~~~~~~~~~~~~~~~~~~~~~
\begin{figure}[!Htb]
\begin{center}
  \includegraphics[scale=.65,angle=0]{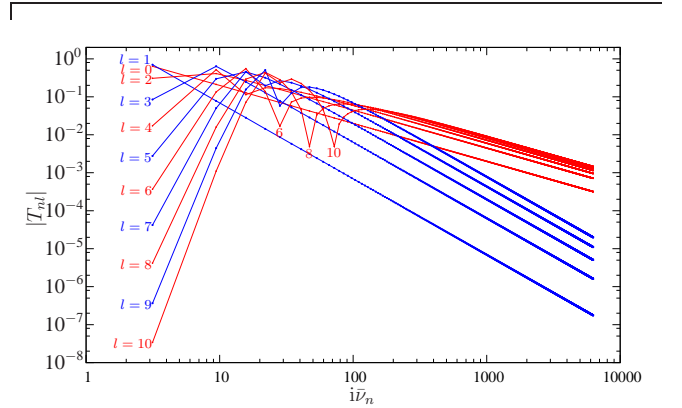}
\end{center}
\caption{(Color online) $|T_{n l}|$ for the first even (red) and odd (blue) Legendre coefficients. The high-frequency tail is reproduced correctly by $t^{(p)}_l$.}
\label{fig:Tnuk}
\end{figure}
%~~~~~~~~~~~~~~~~~~~~~~~~~~~~~~~~~~~~~~~~~~~~~~~~~~~~~~~~~~~~~~~~~~~~~~~~~~~~~~~

%
The transformation matrix from the Legendre to the Matsubara representation is 
\begin{equation}
  T_{n l} \equiv  \frac{\sqrt{2l+1}}{\beta}\int_0^\beta\d\tau \myexp{\i\nu_n\tau}P_l\bigl(x(\tau)\bigr)
\end{equation}
where $\nu_n$ is a fermionic Matsubara frequency and $l$ the Legendre index.
Using (\ref{FourierTransformLegendreAsBessel}) and introducing the reduced frequencies $\bar\nu_n=\beta\nu_n=(2n+1)\pi$, we find
\begin{equation}
  T_{nl}=  (-1)^n\,\i^{l+1} \sqrt{2l+1}\,\, j_l\left(\frac{\bar\nu_n}{2}\right).
\end{equation}
Note that $T_{nl}$ is actually independent of $\beta$.

$T_{nl}$ is a unitary transformation, as can be check explicitely using the Poisson summation formula and the orthogonality of the Legendre Polynomials \eqref{App.A.LegendreOrthogonality}
\begin{align}
  \nonumber
  \sum_{n \in \mathbb{Z}} T^\ast_{nl} T_{nl'} =& \frac{\sqrt{2l+1}\sqrt{2l'+1}}{\beta}
  \times \\
  \nonumber
   \iint_0^\beta d \tau d\tau' & \  P_l(x(\tau)) P_{l'}(x(\tau'))
  \underbrace{  \frac{1}{\beta} \sum_{n\in\mathbb{Z}} e^{-\i \nu_n (\tau - \tau') }} _{= \delta (\tau-\tau')}
\\
  \nonumber
=&   \sqrt{2l+1}\sqrt{2l'+1} \int_{-1}^1 \frac{dx}{2} \ P_l(x) P_{l'}(x)
\\ =& \delta_{ll'}
\end{align}

We will now deduce the coefficients $t^{(p)}_l$ of the expansion of $T_{nl}$
\begin{equation}
  T_{n l} =  \sum_{p\geq 1} \frac{t^{(p)}_{l}}{(\i\bar\nu_n)^p}.
  \label{App.DefSmallt}
\end{equation}
This straightforwardly done from an corresponding representation of the Bessel function, cf. e.g. 
Ref.~\onlinecite{AbramowitzBook}, Section 10.1
\begin{widetext}
\begin{equation}
  j_{l}(z) = z^{-1}\,\times
  \left \{\sin(z-\pi l/2)\sum_{k=0}^{\lfloor \frac{l}{2}\rfloor}(-1)^k\frac{(l+2k)!(2z)^{-2k}}{(2k)!(l-2k)!}\right.
\left.+\cos(z-\pi l/2)\sum_{k=0}^{\lfloor \frac{l-1}{2}\rfloor}(-1)^k\frac{(l+2k+1)!(2z)^{-2k-1}}{(2k+1)!(l-2k-1)!}\right\}.
  \label{App.LargezExpansionOfBessel}
\end{equation}
For the case at hand this gives
\begin{equation}
T_{nl}=-i^l2\sqrt{2l+1}
\left\{\cos\left(\frac{l}{2}\pi\right)\sum_{k=0}^{\lfloor\frac{l}{2}\rfloor}\frac{(l+2k)!}{(2k)!(l-2k)!}\frac{1}{(\i\bar\nu_n)^{2k+1}}\right.
+\i\left.\sin\left(\frac{l}{2}\pi\right)\sum_{k=0}^{\lfloor\frac{l-1}{2}\rfloor}\frac{(l+2k+1)!}{(2k+1)!(l-2k-1)!}\frac{1}{(\i\bar\nu_n)^{2k+2}}\right\}.
\end{equation}
\end{widetext}
The two sums can be combined to
\begin{equation}
T_{nl}=2\sqrt{2l+1}\sum_{p=1}^{l+1}\frac{(l+p-1)!}{(p-1)!(l-p+1)!}\frac{(-1)^p}{(\i\bar\nu_n)^p}\delta_{p+l,\mathrm{odd}},
\end{equation}
which immediately provides the coefficients $t^{(p)}_{l}$ of \eqref{App.DefSmallt}
\begin{equation}
  t^{(p)}_{l} = (-1)^p 2 \sqrt{2l+1} \frac{(l+p-1)!}{(p-1)!(l-p+1)!} \delta_{p+l,\mathrm{odd}}.
\label{eq:smallt}
\end{equation}
Fig.~\ref{fig:Tnuk} shows $T_{n l}$ for the first Legendre coefficients
plotted against the fermionic Matsubara frequency $\i\bar\nu_n$. The doubly
logarithmic plot clearly shows the high-frequency $1/\i\bar\nu_n$-behavior for the
even and the $1/(\i\bar\nu_n)^2$-behavior for the odd coefficients. One can see
that, as expected, structure at very high frequencies is only carried by
polynomials with large values of $l$. 

\end{appendix}
\bibliography{bibextra}

\end{document}